%% file: main.tex
\documentclass[conference]{IEEEtran}
\IEEEoverridecommandlockouts
\usepackage{cite}
\usepackage{amsmath}
\usepackage{amssymb}
\usepackage{amsfonts}
\usepackage{algorithmic}
\usepackage{graphicx}
\usepackage{textcomp}
\usepackage{xcolor}
\def\BibTeX{{\rm B\kern-.05em{\sc i\kern-.025em b}\kern-.08em
    T\kern-.1667em\lower.7ex\hbox{E}\kern-.125emX}}

\usepackage[T1]{fontenc}
\usepackage[utf8]{inputenc}
\usepackage{colortbl}
\usepackage{makecell}
\usepackage{framed}
\usepackage{makecell}
\usepackage{fancybox}
\usepackage{float}
\usepackage{listings}
\usepackage{inconsolata}
\usepackage{subcaption}
\usepackage[hidelinks]{hyperref}
\usepackage{cleveref}
\usepackage{url}
\usepackage{framed}
\usepackage{arydshln}
\usepackage{tcolorbox}
\usepackage{courier} 
\lstdefinestyle{pythonStyle}{
    language=Python,
    basicstyle=\ttfamily\footnotesize,
    commentstyle=\color{green!50!black},
    keywordstyle=\color{violet},
    numberstyle=\tiny\color{gray},
    stringstyle=\color{blue},
    showstringspaces=false,
    emph={self},
    emphstyle=\color{blue},
    breaklines=true,
    breakatwhitespace=true,
    tabsize=2,
    frame=single,
    framesep=2pt,
    xleftmargin=0.0pt,
    xrightmargin=0.0pt,
    rulecolor=\color{black!40},
    backgroundcolor=\color{blue!5},
    morekeywords={True, False, None, __init__},
    captionpos=b,
    escapeinside={*'}{'*},
    extendedchars=true,
    inputencoding=utf8,
    literate={á}{{\'a}}1 {ã}{{\~a}}1 {é}{{\'e}}1,
    columns=fullflexible,
}

\lstdefinestyle{pythonStyle2}{
	language=Python,
	basicstyle=\ttfamily\footnotesize,
	commentstyle=\color{green!50!black},
	keywordstyle=\color{violet},
	numberstyle=\tiny\color{gray},
	stringstyle=\color{blue},
	showstringspaces=false,
	emph={self},
	emphstyle=\color{blue},
	breaklines=true,
	breakatwhitespace=true,
	tabsize=1,
	frame=single,
	framesep=2pt,
	frameround=tttt,
	xleftmargin=2pt,
	xrightmargin=2pt,
	rulecolor=\color{black},
	backgroundcolor=\color{blue!5},
	morekeywords={True, False, None, __init__},
	captionpos=b,
	escapeinside={*'}{'*},
	extendedchars=true,
	inputencoding=utf8,
	columns=fullflexible,
	escapeinside={@@}
}

\definecolor{codegreen}{rgb}{0,0.6,0}
\definecolor{codegray}{rgb}{0.5,0.5,0.5}
\definecolor{codepurple}{rgb}{0.58,0,0.82}
\definecolor{backcolour}{rgb}{0.95,0.95,0.92}
\definecolor{linecolour}{rgb}{0.7,0.11,0.11}

\newcommand{\numQuestions}{60}
\newcommand{\bmk}{Turbulence}
\newcommand{\CCS}{\mathit{CorrSc}}

\newcommand{\com}{\emph{Command}}
\newcommand{\gthree}{GPT-3.5-turbo}
\newcommand{\gfour}{\emph{GPT-4}}
\newcommand{\llamaSeven}{CodeLlama:7B:4-bit-quantised}
\newcommand{\llamaThirteen}{CodeLlama:13B:4-bit-quantised}
\newcommand{\ourpara}[1]{\smallskip\noindent\textbf{#1.}}

\newcommand{\gt}{\emph{GPT-3.5}}
\newcommand{\clsev}{\emph{CodeLlama-7}}
\newcommand{\clthir}{\emph{CodeLlama-13}}
\newcommand{\tzero}{$t\!=\!0$}
\newcommand{\td}{$t\mkern-5mu=\mkern-5muD$}
\newcommand{\pc}[1]{\lstinline|#1|}
\newcommand{\tdtable}{{\textbf{\textit{t $=$ D}}}}
\newcommand{\tzerotable}{{\textbf{\textit{t $\mathbf{= 0}$}}}}

\newtheorem{definition}{Definition}

\def\BibTeX{{\rm B\kern-.05em{\sc i\kern-.025em b}\kern-.08em
    T\kern-.1667em\lower.7ex\hbox{E}\kern-.125emX}}
\begin{document}

\title{Turbulence: Systematically and Automatically Testing
	Instruction-Tuned Large Language Models for Code\\
	\thanks{This work was supported by UK Research and Innovation [grant number EP/S023356/1], in the UKRI Centre for Doctoral Training in Safe and Trusted Artificial Intelligence (www.safeandtrustedai.org).}
}

\author{\IEEEauthorblockN{Shahin Honarvar}
	\IEEEauthorblockA{\textit{Department of Computing} \\
		\textit{Imperial College London}\\
		London, UK \\
	s.honarvar21@imperial.ac.uk}
	\and
	\IEEEauthorblockN{Mark van der Wilk}
	\IEEEauthorblockA{\textit{Department of Computer Science} \\
		\textit{University of Oxford}\\
		Oxford, UK \\
		mark.vdwilk@cs.ox.ac.uk}
	\and
	\IEEEauthorblockN{Alastair F.~Donaldson}
	\IEEEauthorblockA{\textit{Department of Computing} \\
		\textit{Imperial College London}\\
		London, UK \\
		alastair.donaldson@imperial.ac.uk}
}

\maketitle

\begin{abstract}
We present a method for systematically evaluating the correctness and robustness of instruction-tuned large language models (LLMs) for code generation via a new benchmark, \bmk{}.
\bmk{} consists of a large set of natural language \emph{question templates}, each of which is a programming problem, parameterised so that it can be asked in many different forms. 
Each question template has an associated \emph{test oracle} that judges whether a code solution returned by an LLM is correct.
Thus, from a single question template, it is possible to ask an LLM a \emph{neighbourhood} of very similar programming questions, and assess the correctness of the result returned for each question. 
This allows gaps in an LLM's code generation abilities to be identified, including \emph{anomalies} where the LLM correctly solves \emph{almost all} questions in a neighbourhood but fails for particular parameter instantiations.
We present experiments against five LLMs from OpenAI, Cohere and Meta, each at two temperature configurations.
Our findings show that, across the board, \bmk{} is able to reveal gaps in LLM reasoning ability.
This goes beyond merely highlighting that LLMs sometimes produce wrong code (which is no surprise): by systematically identifying cases where LLMs are able to solve some problems in a neighbourhood but do not manage to generalise to solve the whole neighbourhood, our method is effective at highlighting \emph{robustness} issues. 
We present data and examples that shed light on the kinds of mistakes that LLMs make when they return incorrect code results.
\end{abstract}


\section{Introduction}
Large language models (LLMs) have proven effective in tasks such as translating between programming languages \cite{weisz2021} and answering programming questions \cite{lomshakov2023}. 
Their effectiveness has been increased via \textit{instruction tuning}~\cite{LinfengDong2023}, which uses supervision to teach a pre-trained LLM to follow particular kinds of instructions and apply this capability to unseen tasks~\cite{JasonWei2022}.
However, current instruction-tuned LLMs often generate \emph{incorrect} code~\cite{chen2021,hendrycks2021,nguyen2022,liu2023}.
To further enable AI-based code generation in mainstream software development, where correctness and robustness are essential,
it is important to address the issue of developers' lack of trust in LLMs with respect to code generation tasks~\cite{ross2023,perry2022}.
To this end, several works have focused on assessing the ability of LLMs to generate correct code~\cite{chen2021,hendrycks2021,austin2021,xu2022,Yupeng_Chang2024,Zhiqiang_Yuan2023}, whereas other studies have investigated \emph{robustness} while preserving the semantics of prompts or code in their evaluations~\cite{mastropaolo2023,yang2022,tianCJ2023,wang2022,doderlein2022,shirafuji2023,guangYang2023,mingYan2023,anand2021,yueZhuo2024,zhou_yang2024}.
Our paper complements this body of work by introducing a novel dimension to correctness and robustness testing. 
While existing studies primarily assess how models handle semantically equivalent inputs---ensuring that variations do not alter the meaning or functionality of the code or descriptions---we explore how models perform when faced with neighbourhoods of similar but \emph{non-equivalent} tasks.

\ourpara{Our contribution}
Inspired by Gardner et al.~\cite{gardner2020}, the key idea behind our approach is that instead of evaluating an LLM using separate, isolated coding problems, we use sets of related problems, where all problems in a set are variations on a theme---they are all in the same \emph{neighbourhood}.
Rather than being interested in whether an LLM can solve any \emph{particular} problem,
we are interested in identifying \emph{discontinuities} in the LLM's ability to solve a neighbourhood of problems---e.g.\ cases where the LLM correctly solves most problems in a neighbourhood but fails for certain cases.
As opposed to merely identifying problems with isolated code generation prompts (the fact that problematic cases exist is no surprise), identifying discontinuities within a neighbourhood reveals the limits of an LLM’s (in)ability to generalise.

\begin{figure}[t]
	\setlength{\fboxsep}{1pt}
	\begin{tcolorbox}[colback=blue!15, arc=1mm, boxrule=0.5pt, left=0pt, right=0pt, top=0pt, bottom=-10.0pt]
		\captionsetup[subfigure]{aboveskip=-2pt, belowskip=10pt}
		\begin{subfigure}{\columnwidth}
			\begin{tcolorbox}[boxrule=0.3pt, boxsep=2pt,left=2pt,right=2pt,top=2pt,bottom=2pt]
				Write a function called `sum\_even\_ints\_inclusive' that takes one argument, a list of integers, 
				and returns the sum of all even integers from index \boldmath$p_1$ to index \boldmath$p_2$, both inclusive. 
				If no even integers exist in the specified range, the function should return 0.
			\end{tcolorbox}
			\caption{A question template featuring two parameters $p_1$ and $p_2$.}
			\label{fig:Question_Template}
		\end{subfigure}
		\hfill 
		\captionsetup[subfigure]{aboveskip=-2pt, belowskip=10pt}
		\begin{subfigure}{\columnwidth}
			\begin{tcolorbox}[boxrule=0.3pt, boxsep=2pt,left=2pt,right=2pt,top=2pt,bottom=2pt]
				Write a function called `sum\_even\_ints\_inclusive' that takes one argument, a list of integers, 
				and returns the sum of all even integers from index \textbf{1} to index \textbf{8}, both inclusive. 
				If no even integers exist in the specified range, the function should return 0.
			\end{tcolorbox}
			\caption{A question instance from (a) with $p_1=1$ and $p_2=8$.}
			\label{fig:Question_Instance}
		\end{subfigure}
		\hfill
		\captionsetup[subfigure]{aboveskip=-2pt, belowskip=10pt}
		\begin{subfigure}{\columnwidth}
			\begin{lstlisting}[style=pythonstyle2]
def test_odd_range():
  odd_list = [i for i in range(-10001, @$\boldsymbol{p_2}$@*10, 2)]
  assert sum_even_ints_inclusive(odd_list) == 0
			\end{lstlisting}
			\caption{A test case template for (a) featuring $p_2$.}
			\label{fig:Test_Template}
		\end{subfigure}
		\hfill
		\begin{subfigure}{\columnwidth}
			\begin{lstlisting}[style=pythonstyle2]
def test_odd_range():
  odd_list = [i for i in range(-10001, @$\boldsymbol{8}$@*10, 2)]
  assert sum_even_ints_inclusive(odd_list) == 0
			\end{lstlisting}
			\caption{A test case instance from (c) with $p_2=8$.}
			\label{fig:Test_Instance}
		\end{subfigure}
			\captionsetup[subfigure]{aboveskip=-2pt, belowskip=10pt}
		\begin{subfigure}{\columnwidth}
			\begin{lstlisting}[style=pythonstyle2]
def sum_even_ints_inclusive(lst):
  lst = lst[@$\boldsymbol{p_1}$@ : @$\boldsymbol{p_2}$@ + 1]
  return sum([i for i in lst if i % 2 == 0])
			\end{lstlisting}
			\caption{Model solution template for (a) featuring $p_1$ and $p_2$.}
			\label{fig:Solution_Template}
		\end{subfigure}
		\hfill
		\begin{subfigure}{\columnwidth}
			\begin{lstlisting}[style=pythonstyle2]
def sum_even_ints_inclusive(lst):
  lst = lst[@$\boldsymbol{1}$@ : @$\boldsymbol{8}$@ + 1]
  return sum([i for i in lst if i % 2 == 0])
			\end{lstlisting}
			\caption{A model solution instance from (e) with $p_1=1$ and $p_2=8$.}
			\label{fig:Solution_Instance}
		\end{subfigure}
			\hfill
	\end{tcolorbox}
	\caption{An example of a question template, test case template and model solution template, and an instantiation of each}
	\label{fig:Question_Template_Instance_Snippets}
\end{figure}

Our approach is based on the notion of a \emph{question template}.
A question template is a natural language programming specification parameterised by one or more values. 
An example is shown in \Cref{fig:Question_Template}. 
This question template is parameterised by two integer values, $p_1$ and $p_2$, and can be instantiated for any $0 \leq p_1 \leq p_2 \leq K$, where $K$ is a reasonable upper limit for Python list sizes.
An instantiation of the question template of \Cref{fig:Question_Template} with $p_1=1$ and $p_2=8$ is shown in \Cref{fig:Question_Instance}. This is called a \emph{question instance}.

Each question template is paired with an associated \emph{oracle template}.
This includes a suite of parameterised unit tests, featuring the same parameters that appear in the question template.
\Cref{fig:Test_Template} shows an example parameterised test case for the question template of \Cref{fig:Question_Template}.
The parameterised test suite can be instantiated to yield a set of concrete tests for a question instance.
For example, \Cref{fig:Test_Instance} shows the concrete test case obtained by instantiating the test case of \Cref{fig:Test_Template} with $p_1=1$ and $p_2=8$ (as $p_1$ does not occur in the test case template its value is irrelevant to this instantiation).
This test is suitable for checking the correctness of solutions to the question instance of \Cref{fig:Question_Instance}.
An oracle template also includes a \emph{model solution}, which we discuss in \Cref{sec:our_approach}.

Given a (question template, oracle template) pair, an LLM can be asked, via multiple independent queries, to solve a neighbourhood of e.g.\ 100 different question instances derived from the question template, each of which can be automatically checked for correctness via the corresponding instantiated oracle.
The results might be extreme, suggesting that the LLM is completely incapable of solving this neighbourhood of questions (if every solution fails the oracle), or that the LLM can easily solve this neighbourhood of questions (if all solutions pass).
More intriguingly, the LLM might successfully solve many instances of a question template, but unexpectedly yield an incorrect solution for specific parameter values. 
Conversely, it may exhibit a lack of success in addressing the majority of instances in a neighbourhood, yet unexpectedly deliver a correct solution for certain parameter values.
Our method for identifying these discontinuities may offer valuable insights into the limitations of the LLM's reasoning capabilities and has the potential to serve as a source of data for training or fine-tuning.
Furthermore, our approach may feed into discussions as to whether LLMs are truly exhibiting \emph{emergent} reasoning powers, as some researchers have speculated~\cite{saparov2023,shi2023,kojima2022}.
It seems implausible that an LLM that can truly reason would be capable of solving the programming question of \Cref{fig:Question_Template} for many values of $p_1$ and $p_2$ but not, say, for the particular case of $p_1\!=\!100$ and $p_2\!=\!200$.
Prior methods for testing LLM-based code generation using stand-alone problems (see \Cref{sec:related_work}) cannot yield such insights:
key to our method is that it assesses both LLM correctness (by testing each generated code response) and LLM robustness (by assessing how correctness varies across a neighbourhood).

\ourpara{The \bmk{} benchmark}
Conceptually, the method we propose is both LLM- and programming language-agnostic.
We have put it into practice by building a new benchmark,
\bmk{}, for assessing the capability of instruction-tuned LLMs at generating Python code.
\bmk{} comprises (1) infrastructure for automatically assessing LLMs against a set of question and oracle templates, and (2) a set of \numQuestions{} question and oracle templates that we have curated. 
We expect the long-lasting impact of our work to come from (1),
because
our method and infrastructure can be used with any suitable set of question and oracle templates in the future.
Our curated question templates allow us to report results across various state-of-the-art LLMs.
The questions were created from scratch by the paper's authors to avoid direct similarities to existing online questions or code, thus preventing training bias. They were refined based on feedback from a number of experienced Python programmers to minimise any potential ambiguity.

\ourpara{Research questions and summary of findings}
We have used \bmk{} to evaluate a variety of LLMs:
the \gfour{}~\cite{openai_report2023} and \textit{\gthree{}}~\cite{gpt3_5_2023} models from OpenAI \cite{OpenAI2023}, the \com{} model~\cite{CohereCommand2023} from Cohere \cite{CoherePlatform2023}, the 4-bit quantised version of \textit{CodeLlama:7B}, and the 4-bit quantised version of \textit{CodeLlama:13B} with the full precision models and the 4-bit quantised versions being provided by Meta \cite{meta} and Ollama~\cite{ollama}, respectively.
Our evaluation is guided by the following research questions about the instruction-tuned LLMs:

\begin{itemize}
    \item \textbf{RQ1:} How robust are LLMs in code generation when confronted with alterations in fixed values such as numerical values or string characters within prompts?

    \item \textbf{RQ2:} How does setting an LLM's temperature to zero for maximum determinism affect its performance compared to the default temperature?

    \item \textbf{RQ3:} What are the primary errors in the code responses of the LLMs that render the responses incorrect?
    
\end{itemize}

Our findings show that \gfour{} outperformed other LLMs.
Nevertheless, all LLMs exhibited a lack of robustness when faced with variations in many questions. 
Certain question neighbourhoods posed challenges that were either entirely solvable or entirely unsolvable for the LLMs. However, a significant portion of the question neighbourhoods were only partially solved by the LLMs. 
Lowering the temperature reduced the number of partially solved question neighbourhoods, with more falling into either the fully solved or unsolved categories.
This makes sense, because at a lower temperature an LLM will behave more deterministically, so that it is more likely to consistently fail or consistently succeed at a task, whereas at a higher temperature the LLM may behave in a more creative manner, leading to more fluctuation in its ability to solve a given task successfully.
Despite the stochastic nature inherent in LLMs, the partial resolution of some question neighbourhoods \textit{could potentially} highlight gaps in the training data used for the LLMs or flaws in their reasoning.
We present an analysis of the common problems associated with incorrect code generated by LLMs in \Cref{sec:failure_causes}. 

In summary, the main contributions of this paper are:
\begin{itemize}

\item A new approach to assessing correctness and robustness of the code generation capabilities of instruction-tuned LLMs via \emph{neighbourhoods} of related problem instances.

\item \bmk{}, a benchmark and automated testing framework based on our approach, for assessing the Python code generation capabilities of instruction-tuned LLMs.

\item A study using \bmk{} to evaluate the correctness and robustness of five state-of-the-art instruction-tuned LLMs of varying sizes and a deep dive into the key sources of errors in incorrect solutions.
\end{itemize}

In the rest of the paper, we give an overview of our approach (\Cref{sec:our_approach}),
present the \bmk{} benchmark (\Cref{sec:curating_questions}), present results applying \bmk{} to a range of instruction-tuned LLMs (\Cref{sec:experimental}), and discuss characteristics of incorrect code returned by LLMs (\Cref{sec:failure_causes}).
We discuss threats to validity (\Cref{sec:threats}) and related work (\Cref{sec:related_work}) before concluding (\Cref{sec:conclusion}).

\section{Our Benchmarking Approach}\label{sec:our_approach}
\begin{figure}[t]
    \centering
\includegraphics[width=252pt]{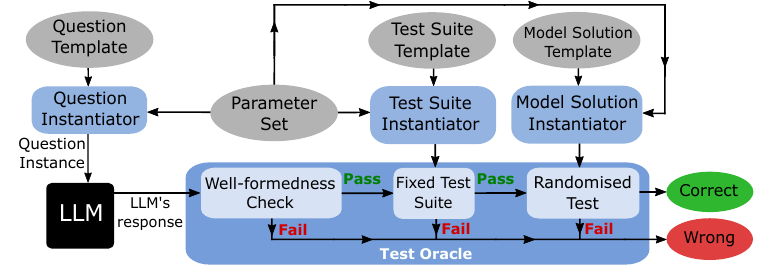}
    \caption{Overview of our benchmarking approach}\label{fig:architecture}
\end{figure}

We now describe our general approach to benchmarking LLMs for code, an overview of which is shown in \Cref{fig:architecture}.
In \Cref{sec:curating_questions} we describe \bmk{}, a concrete benchmark based on this approach, tailored towards testing LLMs concerning Python code generation.
However, our approach is LLM- and programming language-agnostic, allowing future testing of other LLMs across various programming languages.

\ourpara{Question Templates and Instances}
A \emph{question template} is a programming problem expressed in natural language, featuring one or more parameters. Recall the template of \Cref{fig:Question_Template}, which takes integer parameters $p_1$ and $p_2$.
Instantiating this template with $p_1=1$ and $p_2=8$ yields the \emph{question instance} of \Cref{fig:Question_Instance} (instantiated parameters are shown in bold for clarity).
Each question template should be accompanied by a \emph{parameter set}: a set of suitable parameter valuations that are meaningful for the question template.
The \bmk{} benchmark described in \Cref{sec:curating_questions} is equipped with a generator that automatically produces a suitable parameter set of a desired size for a given question template.
A question template can be automatically instantiated with a range of parameter values drawn from its parameter set, leading to a set of question instances that can be presented to an LLM (see \Cref{fig:architecture}).

Intuitively, since question instances from the same template differ only by parameter values, they should all be just as easy or hard to solve as each other. Thus, it is noteworthy when an LLM solves some but not all instances from a template.

\ourpara{Assessing Correctness: Oracle Templates}
To assess whether an LLM has returned a \emph{correct} solution to a question instance, the benchmark designer must provide an \emph{oracle template} for each question instance.
This comprises: (1) a \emph{fixed test suite}---a set of unit tests, parameterised with the same parameters as the question template, which when instantiated provides a concrete test suite for the question instance;
(2) a \emph{model solution template}, which can be instantiated to provide a correct solution for any question instance;
and (3) a \emph{random input generator}, which facilitates fuzz testing of solutions as described further below.

To illustrate this, consider the question template of \Cref{fig:Question_Template}.
The oracle template associated with this question template comprises multiple \emph{parameterised} test cases. One of these is shown in \Cref{fig:Test_Template}, and refers to parameter $p_2$ from the question template.
When the question template is instantiated with $p_1=1$ and $p_2=8$, as shown in \Cref{fig:Question_Instance}, the oracle template is also instantiated with these parameters. This transforms the parameterised test case of \Cref{fig:Test_Template} into the \emph{concrete} test case of \Cref{fig:Test_Instance}, which is suitable for assessing the correctness of a solution to the concrete question instance of \Cref{fig:Question_Instance}.
Furthermore, \Cref{fig:Solution_Template} shows a parameterised model solution for the question template, again expressed in terms of the parameters $p_1$ and $p_2$, while \Cref{fig:Solution_Instance} shows a concrete instantiation of this model solution for the given parameter values. This concrete model solution facilitates experimental comparison with an LLM-generated solution on arbitrary input values.
The random input generator component of the oracle template (not shown in \Cref{fig:Question_Template_Instance_Snippets}) provides a means of generating a stream of input values at random to support this kind of comparison.

Armed with these components, a code solution returned by an LLM in response to a question instance is deemed \emph{correct} if and only if all of the following hold (see the ``Test Oracle'' component of \Cref{fig:architecture}):
the LLM solution is \emph{well-formed} (syntactically correct and conforming to any static typing rules of the programming language);
the LLM solution passes all tests in the \emph{fixed test suite} (instantiated with the parameters associated with the question instance);
and the LLM solution yields the same result as the \emph{model solution} (again, instantiated with the parameters of the question instance) when applied to a number of random inputs generated by the input generator.
The approach of comparing the LLM solution with a model solution using randomly-generated inputs is a special case of fuzzing known as \emph{random differential testing}~\cite{McKeeman}.
The combination of testing via a fixed test suite and through random differential testing helps to ensure that the LLM solution works on particular important edge cases (provided by the fixed tests), as well as on a wider range of examples (from the randomised input generator).
The user can control the amount of randomised testing per question instance.

\ourpara{Avoiding Ambiguity} It would be unfair to penalise an LLM for failing test cases that check aspects of a question whose solution is open to multiple interpretations.
The designer of a question and oracle template must either (a) state the question precisely, without ambiguity, or (b) design the oracle template to avoid testing solutions in ambiguous parts of the input space. For example, the question template in \Cref{fig:Question_Template} avoids ambiguity by specifying that list indices are \emph{inclusive}. Alternatively, this clarification could be omitted, and the oracle template could be adjusted to exclude test cases with even integers at indices $p_1$ and $p_2$, ensuring the oracle does not distinguish between solutions treating index ranges as inclusive or exclusive.

In \Cref{sec:curating_questions} we explain how we used feedback from human programmers to avoid ambiguity in \bmk{}.

\ourpara{Assigning Correctness Score}
An oracle template provides a means for assigning a pass/fail result to an LLM's solution for a question instance. 
We explain how these results are combined into an overall score for each question template, reflecting the LLM's effectiveness in solving that question neighbourhood. Given the non-deterministic nature of LLMs, multiple independent queries per question instance are necessary.

\begin{definition}[Correctness Score]\label{def:CCS}
Let $L$ be an LLM under evaluation.
Let $Q$ be a question template with $M$ associated parameter valuations (so that $M$ distinct question instances are derived from $Q$).
Suppose that the LLM is queried $N$ times per question instance, and let $L_i^j(Q)$ denote the result returned by $L$ the $j$th time that it is queried with question instance $i$ of $Q$. 
Let $\mathit{Oracle}(L_i^j(Q))\!=\!1$ if this result is deemed correct according to the oracle template and 0 otherwise.
The \emph{correctness score}, $\CCS{}$, for question template $Q$ with respect to LLM $L$, $\CCS{}(Q, L)$, is then defined as follows:
  
\begin{equation*}\label{eq:correctness}
    \CCS{}(Q, L) = ~ \frac{\sum_{i=1}^{M}\sum_{j=1}^{N}\mathit{Oracle}(L_i^j(Q))}{M\times N}
\end{equation*}
\end{definition}

This is the mean over the correctness of all solutions returned by the LLM, where an individual solution is given a score of either 0 or 1.
It yields a score in the range $[0, 1]$ for each question template $Q$.
Rather than featuring $M \times N$ discrete question instances, our approach entails the usage of $N$ identical collections of $M$ unique question instances. 
This design choice is motivated by the inherent non-deterministic nature of LLMs, allowing multiple attempts with the same question instance to assess the LLM's ability to generate correct responses.

While the $pass@k$ metric \cite{chen2021}, commonly used for evaluating LLMs in code generation, is a well-regarded measure, it is not suitable in the context of our study as our goal was to assess the overall correctness of each model.
The $pass@k$ metric \cite{chen2021} focuses on whether at least one correct solution is found within the first $k$ attempts. 
In our study, each prompt was sent five times ($k=5$). Consider the following hypothetical results for a single prompt:  incorrect (attempt 1), correct (attempt 2), incorrect (attempt 3), correct (attempt 4), and incorrect (attempt 5).
In this case, $pass@5$ would be 100\% since at least one correct answer is provided, but the overall correctness (i.e.\ the proportion of correct answers) is 40\% (or 0.4, as described in \Cref{def:CCS}). We focused on the CorrSc metric to evaluate overall correctness.

\section{The \bmk{} Benchmark}\label{sec:curating_questions}
Based on the approach described in \Cref{sec:our_approach} we have created a novel benchmark, \bmk{}, for evaluating the correctness and robustness of instruction-tuned LLM for code. 
\bmk{} focuses on the generation of Python code due to the language's popularity and the ample Python training data available for LLMs.

To create the benchmark, we developed a diverse set of 60 Python problem-solving questions encompassing fundamental concepts and basic data structures ensuring comprehensive and balanced coverage of key topics. No specific methodology or framework was followed for the selection of these questions, and to the best of our knowledge, no existing research outlines best practices for question formulation in this context. Furthermore, as discussed later, we deliberately avoided reusing publicly available questions to mitigate potential training bias. 

\Cref{table:problem_groups} provides a broad categorisation of the 60 question templates into six distinct problem groups, further subdivided into subgroups. The questions utilise a wide variety of Python data types, either explicitly mentioned in the questions or implicitly required in the solutions. These include list (43 questions), integer (35 questions), boolean (60 questions), string (39 questions), set (9 questions), tuple (4 questions), and NumPy matrix (2 questions). Since some questions span multiple problem groups and data types, the counts in \Cref{table:problem_groups} and the data type totals exceed 60. Additionally, the subgroup counts within each problem group may exceed the total for their respective group, as certain questions pertain to more than one subgroup. 

\bmk{} comprises 60 question templates, each featuring at least one parameter,
where each parameter is either a numerical value or a string.
Each question template is equipped with an associated oracle template: a fixed test suite, random input generator and model solution, as described in \Cref{sec:our_approach}.
Every question template is accompanied by a parameter set of size 100, yielding 100 question instances per template. Hence, a total of 6,000 question instances are generated by \bmk{}. 
For each question template, the parameter set was created by choosing a number of natural or evidently interesting parameter valuations (e.g.\ to exercise edge case behaviour), and thereafter populated with random valuations, restricted to well-formed valuations. For example, with respect to the question template of \Cref{fig:Question_Template} we would not allow negative values or values such that $p_2 < p_1$.

To avoid problems of bias occurring due to LLMs having been exposed to questions during the training, we decided to write question templates ourselves, from scratch, rather than seeking existing questions available on the internet.
This was done to ensure that the LLMs were not able to simply regurgitate previously learned information, but instead had to generate new and creative responses. 
We were also careful not to put our questions online publicly before running experiments against LLMs.
We used only a small selection of the most trivial questions when undertaking preliminary evaluation against LLMs during the construction of \bmk{}, to guard against the possibility of these (closed source) LLMs learning from our interaction with them.

\begin{table}[h!]
\caption{A classification of the \bmk{} questions into problem groups and subgroups}\label{table:problem_groups}
  \input{question_classes}
\end{table}

To ensure the clarity of question templates and the correctness of test oracles, we asked two experienced Python programmers to solve an instance of each question template independently, cross-checking their solution against our test oracle, and soliciting their feedback about potential ambiguity in the question.
This led us to improve the wording of several question templates and fix several bugs in our test oracles.

Creating our own questions has its pros and cons.
As argued above, using previously-unseen questions minimises problems of training-related bias, but it could arguably be more interesting to have a benchmark based on real-world programming challenges faced by developers ``in the field''.
While the true role of LLMs in software engineering is solving real-world programming tasks, to have any chance of being useful in such contexts they should \textit{at least} be capable of solving the kinds of programming problems that beginner to intermediate programmers would be capable of solving.
Also, we emphasise that \bmk{} is just one example of our proposed approach in \Cref{sec:our_approach}.
The enduring value of our research lies in the approach itself, which could be retargeted to use alternative questions.

We deliberately included questions that, while uncommon in typical development scenarios, are pertinent to evaluating the reasoning capabilities of LLMs. For instance, a prompt like \textit{``Write a function called `all\_ints\_exclusive' that takes one argument, a list of integers, and returns the list of all elements from index 0 to index 1, both exclusive''} serves as an edge case designed to test the model's ability to comprehend and execute nuanced instructions. A primary goal of our benchmark is to assess whether LLMs are genuinely exhibiting emergent reasoning abilities. True reasoning capability should enable a model to solve not only standard problems but also edge cases that deviate from common patterns.
While a human developer is unlikely to craft such an edge-case prompt, it is important to consider the evolving contexts in which LLMs are deployed.
LLMs are increasingly being used in the back-ends of systems (such as integrated development environments) where prompts are generated programmatically rather than being written by humans.
In these automated systems, the likelihood of encountering edge cases rises, as the prompts may not undergo human refinement or oversight. Auto-generated prompts are inherently more prone to exhibiting unusual or unexpected parameters, making it essential for LLMs to handle them effectively.
Moreover, evaluating out-of-distribution robustness has been recognised as a critical aspect in the field of NLP: as highlighted by Yuan et al.\ \cite{Yuan23}, assessing how models perform on data that falls outside the distribution of their training data is necessary for understanding their generalisation capabilities and identifying potential weaknesses.

We distinguish between (a) the underlying conceptual framework of \bmk{} and (b) the specific empirical findings of this study. It is evident that (a), the concept of using neighbourhoods to identify reasoning discrepancies in LLMs, could be extended to test other LLMs (with a small amount of engineering effort specific to each model) and could also be adapted for other programming languages (requiring additional engineering effort for each language). However, concerning (b), we do not expect our specific findings to generalise directly to other LLMs or programming languages. Instead, we anticipate discovering different deficiencies, similar to how applying a software testing technique to different systems under test reveals distinct bugs. 
Our findings show that our approach effectively provides valuable insights into code generation.

\ourpara{Practical Issues}
Implementing our approach necessitates some level of prompt engineering \cite{reynolds2021} to enhance the chances of obtaining source code from an LLM. In our initial experiments with the models discussed in \Cref{sec:experimental}, we observed that appending a simple prefix requesting that Python code be enclosed within triple backticks proved effective. The returned code could then be extracted by locating the section between the triple backticks.

\section{Experimental Evaluation}\label{sec:experimental}
We now present results from running \bmk{} against a range of LLMs.

\subsection{Experimental Setup}\label{subsec:experimental_setup}
We have gathered results running \bmk{} against five LLMs: \gfour, \emph{\gthree{}}, \com, \emph{\llamaSeven{}}, and \emph{\llamaThirteen{}}.

\gfour{} \cite{openai_report2023}, by OpenAI \cite{OpenAIabout2023}, is a large multimodal text generation model. \gthree{} is the most advanced in the 3.5 series, trained on text and code up to Q4 2021~\cite{gpt3_5_2023}. Cohere’s 52-billion-parameter Command model \cite{CohereCommand2023} generates text from user commands. Meta’s CodeLlama family \cite{codellama2023} offers coding-specialised models (7B, 13B, 34B, 70B), including instruct-tuned versions. Ollama \cite{ollama} provides quantised versions like \llamaSeven{} and \llamaThirteen{}, which run on standard machines with 4GB and 8GB of RAM. We selected these models due to their low resource demands: \gfour{}, \emph{\gthree{}}, and \com{} are hosted remotely, while \emph{\llamaSeven{}} and \emph{\llamaThirteen{}} are small enough that they can be run locally as described below.

To maintain conciseness, in the rest of this paper, we refer to \textit{CodeLlama:7B:4-bit-quantised}, \textit{CodeLlama:13B:4-bit-quantised}, and \textit{GPT-3.5-turbo} as \clsev{}, \clthir{}, and \gt{}, respectively. \emph{LLM configuration} denotes an LLM combined with a specific temperature setting and \textit{\tzero{}} and \textit{\td{}} refer to the configurations of the LLM with temperature settings of 0 and default, respectively.

We accessed proprietary models \gfour, \gt{}, and \com{} via their commercial APIs. Initially, we minimised \bmk{}-related queries to these models to prevent potential bias, making only a few simple queries to test and debug the \bmk{} infrastructure.

We downloaded \clsev{} and \clthir{} from the Ollama website \cite{ollama} and ran them on a
MacBook Pro with an Apple M1 Pro CPU and 16GB RAM, running macOS 14.0. To prevent bias, our queries never included sample solutions or hints about the correctness of previous responses from the LLM under test.

We evaluated all LLM-generated solutions for correctness on a desktop machine with an Intel Core i7-12700 CPU and 16GB RAM, running Ubuntu 22.04.2.

Every LLM has a user-determined \emph{temperature} parameter that controls output randomness. Lower temperatures reduce randomness, improving quality but decreasing diversity~\cite{caccia2020,hashimoto2019}, while higher temperatures increase randomness, enhancing creativity.
In addressing \textbf{RQ2}, we focused exclusively on comparing the models' behaviour at two specific settings: their default temperature and a temperature of zero. A temperature of zero was chosen to evaluate the performance of LLMs in as deterministic a context as possible, causing the models to select the most probable next token at each step.
The default temperatures varied across LLMs, as developers fine-tuned them for optimal performance, balancing diverse outputs and coherence. Since these default temperatures reflect the intended behaviour envisioned by the developers, we used them to assess the performance of LLMs in their standard operational settings. Our goal was to analyse the shift from non-deterministic behaviour at default temperature to maximum determinism at a temperature of zero, offering clear insights into how determinism affects LLM performance without the complexity of varying randomness. We excluded broader temperature values to keep our experiments tractable.

Due to the stochastic nature of LLMs, repeat runs of experiments are necessary.
At the same time, access to commercial LLMs (i.e.\ \gt{}, \gfour, and \com) is costly, with variable query times.
We ran the full benchmark 5 times for each LLM configuration.

While a temperature of 0 should cause the LLMs to behave in a highly deterministic manner, our initial mock tests revealed that LLMs occasionally yielded varying answers. This non-deterministic behaviour may be due to several factors, including non-deterministic GPU operations, memory access patterns, and numerical precision~\cite{determinism} and the inherent randomness from sampling, even at a temperature of 0~\cite{Gawlikowski2022}.

Our results are thus based on a comprehensive set of 300,000 LLM responses (i.e.\ number of models\,$\times$\,number of configurations per model\,$\times$\,number of prompts per each model's configuration\,$\times$\,number of repeat runs $= 5 \times 2 \times 6000 \times 5 = 300000$). For a consistent comparison of experimental results, we used the same random seed when generating parameters for each question template.

\begin{figure*}
    \centering
    \includegraphics[width=0.85\textwidth]{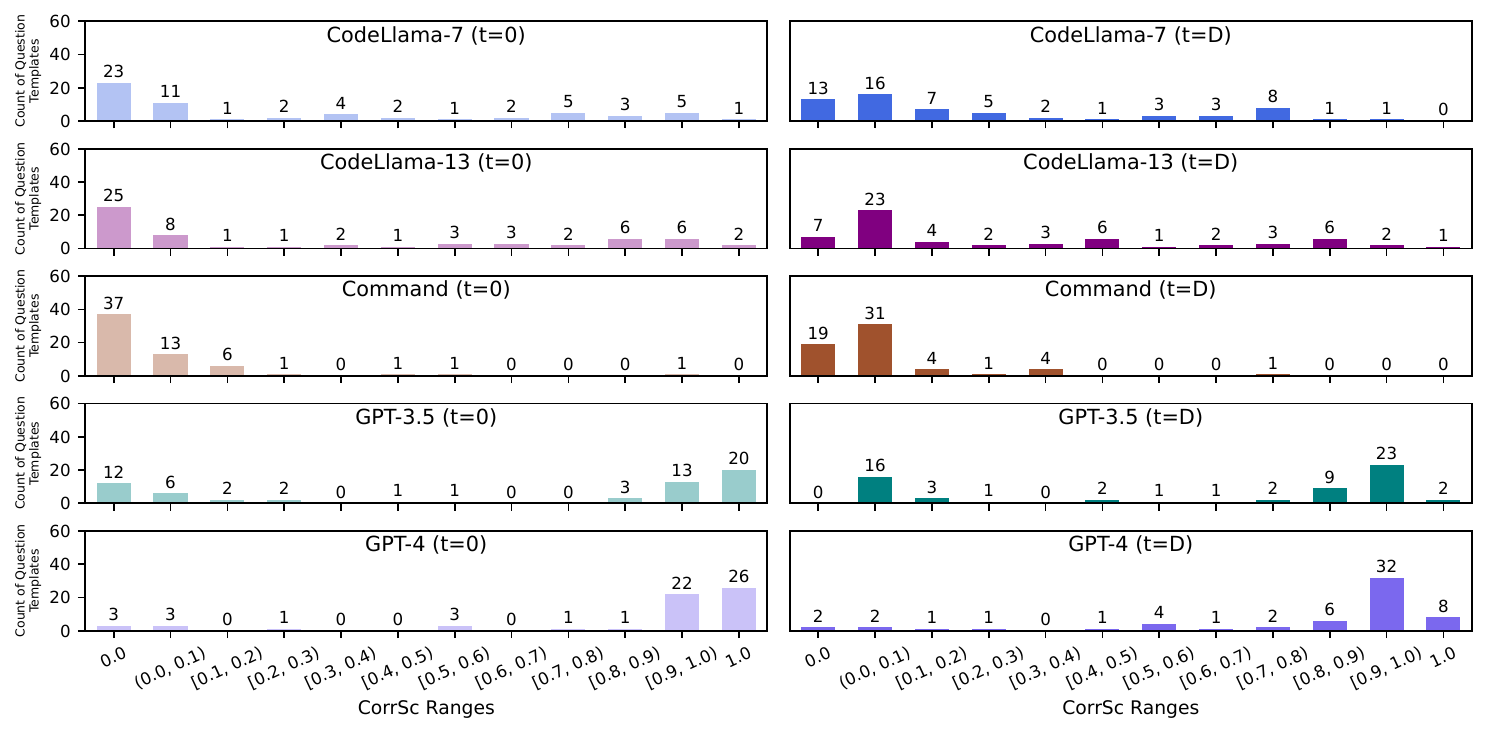}
    \caption{$\CCS$ of question templates across the LLM configurations evaluated}\label{fig:CCS}
\end{figure*}

\begin{figure}
	\begin{tcolorbox}[colback=blue!15, arc=1mm, boxrule=0.5pt, left=0pt, right=0pt, top=0pt, bottom=-5.0pt]
		QT: \textit{{\spaceskip=2pt plus 3pt minus 1pt Write a function called `sum\_of\_multiples'} that takes one argument, a positive integer, and returns the sum of the first \textbf{p} positive multiples of the given integer.}\\
		\begin{subfigure}{0.495\columnwidth}
			\caption{$p=51$, correct answer}
			\begin{lstlisting}[style=pythonstyle2]
def sum_of_multiples(n):
				
  if n <= 0:
    return 0
				
  sum = 0
  for i in range(1,52):
    sum += i * n
				
  return sum
			\end{lstlisting}
			\label{fig:pf_1}
		\end{subfigure}
		\hfill 
		\begin{subfigure}{0.49\columnwidth}
			\caption{$p=56$, wrong answer}
			\begin{lstlisting}[style=pythonstyle2]
def sum_of_multiples(n):
  if n <= 0:
    raise ValueError("n must be a positive integer")
  sum = 0
  for i in range(1, 57):
    if i % n == 0:
      sum += i
  return sum
			\end{lstlisting}
			\label{fig:pf_2}
		\end{subfigure}
	\end{tcolorbox}
	\caption{Consistent failure, \clsev{} (\tzero{})}
	\label{fig:partial_failure}
\end{figure}

\begin{figure*}
    \centering
    \includegraphics[width=0.95\textwidth]{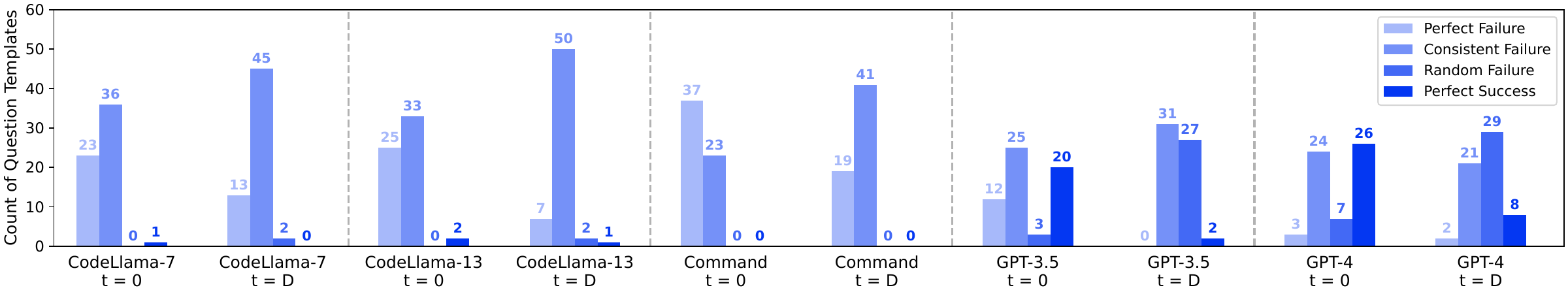}
    \caption{Distribution of \bmk{} question templates across result categories for the LLM configurations evaluated}\label{fig:CCS_Groups}
\end{figure*}

\subsection{Results Based on CorrSc}\label{subsec:results}
\Cref{fig:CCS} summarises how well each LLM configuration solved question templates. Recall from \Cref{def:CCS} that we obtain the correctness score, $\CCS{}$, for each question template and LLM configuration. In \Cref{fig:CCS}, the first and last bars of each graph indicate how many question templates each LLM configuration either completely failed (scoring 0) or perfectly succeeded (scoring 1.0). The intermediate bars represent scores in ranges of size 0.1. For each LLM configuration, the corresponding graph shows the number of question templates whose \textit{CorrSc} fell within each range. The density of question templates is mostly clustered around high or low score ranges, with no abrupt changes in the middle. For \clsev{}, \clthir{}, and \com{} at both temperatures, the data density is mostly in the lower score ranges, indicating weaker performance. In contrast, \gt{} and \gfour{} show superior performance with more templates in the higher score ranges. 
However, there are numerous question templates where the LLMs failed to address all instances. 
Regarding \textbf{RQ1}, the many question templates not scoring 1.0 highlight the limited robustness of LLMs in addressing question neighbourhoods.

Addressing \textbf{RQ2}, examining each vertical pair of plots in \Cref{fig:CCS} shows that with temperature zero, the data distribution shifts to the left side for \clsev{}, \clthir{}, and \com{}; and to both sides for \gt{} and \gfour{} indicating that LLMs are more likely to either fail consistently or succeed consistently for a given question neighbourhood.

\subsection{Results Based on Distinct Categories}\label{subsec:groups}

Recall that each question template is instantiated to yield a neighbourhood of 100 instances, each using distinct parameters, and each instance is given to the LLM across 5 rounds.
We categorise LLM performance for a question into four distinct categories: \textit{perfect failure}, where the LLM does not ever return a correct result ($\CCS{}\!=\!0.0$, i.e.\ the LLM cannot solve this neighbourhood of questions at all); \textit{perfect success}, where the LLM always returns a correct result ($\CCS{}\!=\!1.0$, i.e.\ the LLM can solve this neighbourhood effortlessly); \textit{consistent failure}, where the LLM returns at least one correct result but where there is at least one instance for which the LLM returns an incorrect result across all 5 rounds (the LLM appears to be completely blocked on at least one question instance);
and \textit{random failure}, where the LLM returns at least one incorrect result, but there is no instance for which the LLM returns an incorrect result across all 5 rounds (the LLM is not completely blocked on any question instance). This category describes cases where the LLM \emph{does} appear capable of generalisation, solving every instance in a neighbourhood in at least one round, with sporadic failures due to its stochastic nature, not necessarily a lack of reasoning ability.

The \textit{consistent failure} category is particularly noteworthy as it suggests a reasoning gap---an inability to generalise, since the LLM \emph{can} sometimes solve instances arising from the question template, but there are certain instances that it does not manage to solve on any round. \Cref{fig:partial_failure} shows a \emph{consistent failure} example:
when the question template (referred to as QT in the figure) was instantiated with $p\!=\!51$ (\Cref{fig:pf_1}), \clsev{} at \tzero{} solved the question instance correctly in all five rounds (that is at least one correct answer); however, when the question template was instantiated with 56 (\Cref{fig:pf_2}), the LLM consistently generated wrong answers in all five rounds.
The \emph{consistent failure} category highlights LLM robustness issues where failures are consistent for particular question instances, and not solely due to the LLM's stochastic nature.
Resource and financial constraints limited our experiments to five rounds; consistent failure across more rounds would offer stronger evidence of LLM reasoning gaps.

\Cref{fig:CCS_Groups} shows results according to these categories.
The \emph{perfect failure} and \emph{perfect success} bars mirror the $\CCS{}=0.0$ and $\CCS{}=1.0$ bars of \Cref{fig:CCS}, respectively.

To answer \textbf{RQ1}, \Cref{fig:CCS_Groups} indicates a lack of robustness in the performance of the LLMs considered in this study. Notably, there is a significant count of question templates categorised as \emph{consistent failure} for each LLM configuration. \clthir{} (\td{}) exhibits the highest count (50 question templates), while~\gfour{} (\td{}) has the smallest count (21 question templates). 

To address \textbf{RQ2}, lowering the temperature from the default value to zero led to a reduction in the number of question templates classified under the \emph{consistent failure} category, with the exception of \gfour{}. A similar trend was observed in the \emph{random failure} category, where all models except \com{} showed a decrease in the number of question templates. This effect was especially noticeable in the \textit{GPT} models.

\section{Exploring Reasons for Failure}\label{sec:failure_causes}
In this section, we address \textbf{RQ3} by examining the main errors in the LLM's code responses that caused them to be incorrect.
\Cref{table:failure_reasons} presents 9 failure categories, which we now illustrate with selected examples. 
The initial three rows in \Cref{table:failure_reasons}, i.e.\ \textit{no function}, \textit{wrong function name}, and \textit{wrong count of arguments}, align with the \textit{well-formedness check} of the \bmk{} test oracle (\Cref{fig:architecture}). 
Recall that each question instance asks the LLM to write a \textit{Python function} bearing the \textit{specified name} and \textit{count of arguments} (\Cref{fig:Question_Instance}).

\begin{table*}
\caption{Percentage of responses per LLM configuration out of 30,000, categorised by test failures and passes}\label{table:failure_reasons}
  \input{failure_reasons}
\end{table*}

For the remaining six categories, the \emph{syntax error} category is identified via the Python parser,
the \emph{static type error} category is identified using the Pylint linter~\cite{Pylint},
and the remaining categories are identified by running the test oracle for a question instance.

\ourpara{No function}
This category includes cases where the LLM failed to generate a Python function, occurring with \com{} (\td{}) and \gt{} (\td{}) for certain questions. 

\ourpara{Wrong function name}
This occurs when the LLM generates a function with a different name than requested, causing failures as each \bmk{} test oracle requires specified function names, and there is no systematic, reliable way to fix function names, especially when responses include multiple functions. At \td{}, all LLMs exhibit this problem in some cases.

\ourpara{Wrong number of arguments} 
Here, the LLM generates a correctly-named function but with the wrong number of arguments, making it incompatible with the test oracle. All LLMs except \gt{} and \gfour{} experienced this problem. 
\Cref{table:failure_reasons} shows this issue was less frequent at \tzero{}.

\ourpara{Syntax errors} Here, the LLM output could not be parsed due to Python syntax errors, such as unmatched parentheses, misaligned brackets, incorrect indentation, missing comment indicators (\texttt{\#}), using \pc{else if} instead of \pc{elif}, missing \pc{except} or \pc{finally} clauses after a \pc{try} block, invalid variable names (e.g.\ \pc{6th\_number}), and invalid assignment targets (e.g.\ \pc{len(binary) -= 1}). Syntax errors decreased when LLMs were run at temperature 0.

\ourpara{Static type errors} This is where the integrated Pylint linter \cite{Pylint} identified static type errors in generated code, including undefined variables (e.g.\ using \pc{math.gcd} without importing the \pc{math} library), and the use of Python keywords as variable names (e.g.\ \pc{sum = sum(multiples)}). Lowering the temperature from default to 0 in all LLMs (except \clthir{} and \com{} models) reduced both the quantity and variety of static type errors in the generated responses.

\ourpara{Resource exhaustion error} This category involves cases where the generated code exceeded time or memory resources during execution, observed across all LLMs, despite more efficient solutions being available. For example, \gfour{} (\td{}) was asked to write a function to return the number of subsets of size 54 from a set of elements. The LLM used \pc{len(list(combinations(elements,54)))}, which becomes resource-intensive for sets larger than 60. A more efficient solution is to use \pc{math.comb(len(elements),54)}, which works efficiently with any size of a set.

\ourpara{Runtime errors}
This category covers cases where executing the generated code caused Python errors. For example, \pc{for i in range(2,x+1)} where \pc{x} was a tuple, making concatenation of an `int' and a `tuple' invalid.

\ourpara{Assertion errors and fuzzing failures} 
This category covers cases where the code executed but was functionally incorrect, not fitting the errors in the first seven rows of \Cref{table:failure_reasons}. We identified five main causes of functional errors and grouped them accordingly. The first group involves incorrect ranges. For questions requiring an index or number range, the common bug was either excluding the range or inaccuracies in the lower or upper bounds. For example, when \clthir{} (\tzero{}) was asked to return all list elements from index 55 to 98, both inclusive, it returned elements from $[54:99]$ instead of the expected range, $[55:99]$.

The second group includes responses where redundant code rendered the functionality incorrect, though the code would have been correct without it. 

\begin{figure}
	\begin{tcolorbox}[colback=blue!15, arc=1mm, boxrule=0.5pt, left=0pt, right=0pt, top=0pt, bottom=-5.0pt]
QI: \textit{Write a function called `prime\_factors' that takes one argument, a list of positive integers, and returns the set of all prime factors of the integer at index \textbf{85}}.
        \begin{lstlisting}[style=pythonstyle2]
def prime_factors(lst):
  number = lst[85]
  factors = set()
  for i in range(2, int(number/2) + 1):
    while number % i == 0:
      factors.add(i)
      number = number/i
  if number > 2:
    factors.add(number)
  return factors
       \end{lstlisting}
       \end{tcolorbox}
    \caption{Functional bug, code generated by \gt{} (\td{})}
    \label{fig:g3_wrong_response}
\end{figure}

The third group includes responses with an incorrect logical order of tasks. For example, when \clsev{} (\td{}) was asked to return the second largest number from a specified index range in a list, its generated code sorted the entire list first, then sliced it and returned the second element. The correct approach is to slice the specified range first, then sort it, and return the second element.

The fourth group includes responses where the LLM misunderstood the question. For example, \com{}~($t\!=\!0$) was asked if the integer at index 85 of a list was a perfect number, it generated the incorrect code: \pc{return nums[85]==64820}, where \pc{nums} was a placeholder for the list.

The fifth group includes partially correct responses that miss certain input cases. In \Cref{fig:g3_wrong_response} (where QI refers to \emph{question instance}), the code fails to return \{2\} when the list contains 2 at index 85 due to the condition \pc{if number > 2} instead of \pc{if number >= 2}, which would correctly identify 2 as a prime number.

\ourpara{Passed} As shown in the ``passed'' row in \Cref{table:failure_reasons}, reducing the temperature to 0 increased the number of correct answers for all models except \com{}. \gfour{}, which outperformed \textit{GPT-3.5}, solved over 82\% of question instances in both settings, while \com{} underperformed across all setups. Additionally, \clthir{} outperformed \clsev{}.

\ourpara{Patterns in parameter values leading to errors}
Across most question templates we could not discern patterns among parameter values that led to incorrect LLM answers.
The values were highly diverse with no identifiable trends. However, when examining cases where the LLM succeeded with most of the parameters in a question neighbourhood but failed with specific parameters, i.e.\ $0.9 \le CorrSc < 1.0$, we identified certain patterns. \clsev{}, \clthir{}, \textit{GPT-3.5}, and \gfour{} generated incorrect answers for question instances involving indices or number ranges, particularly at temperature 0. In contrast, \com{} showed no pattern at any temperature. The pattern involved indices or bounds that were identical, where the lower bound was 0, or where the difference between bounds was 1. In other words, if $x$ denotes a non-negative integer, the ranges were: $(0, x), (x, x), [0, x], [x, x],$ and $(x, x+1)$, with parentheses indicating exclusivity and square brackets indicating inclusivity. Additionally, \textit{GPT-3.5} made errors when inserting a character before another in a string if either was a space character.

\section{Threats to Validity}\label{sec:threats}
Our assessment of LLM correctness relies on (a) unambiguous questions, (b) accurate test oracles that do not mis-classify a correct answer as incorrect, and (c) strong test oracles capable of catching errors. For (a) and (b), although skilled Python programmers reviewed our questions and test oracles (see \Cref{sec:curating_questions}), this may not have eliminated all potential ambiguities and oracle errors. For (c), we combine regression testing and random differential testing to thoroughly evaluate LLM responses, but testing is inherently incomplete.

Our findings are limited to the LLMs we evaluated; however, \bmk{} supports integrating additional LLMs in the future. To balance costs, we set $M$ and $N$ (\Cref{def:CCS}) to 100 parameter settings per question for diversity and 5 runs to address variance, reflecting our resource constraints. Larger values would be preferable with increased resources.

To avoid training data bias, we developed custom question templates instead of using internet-sourced ``real-world'' questions. While our questions may resemble online ones, they are tailored for the question neighbourhood approach. In practical tasks, parameters in \bmk{} questions are usually kept as formal parameters, awaiting user input. However, artificial questions in \bmk{} offer key advantages: (i) focusing on hard-to-find edge cases and (ii) enabling easy reproducibility and comparison across LLMs. Our framework remains flexible for future studies with different question sets.

\section{Related Work}\label{sec:related_work}
We categorise prior work on LLM robustness and correctness for code into the following themes.

\ourpara{Correctness and evaluation benchmarks} Various benchmarks and datasets, such
as HumanEval~\cite{chen2021}, APPS~\cite{hendrycks2021}, MBPP~\cite{austin2021}, CodeContests~\cite{YujiaLi2022}, CodeXGLUE~\cite{lu2021}, and examples from LeetCode~\cite{nguyen2022} have been used to evaluate the correctness of LLM-generated code,
and the effectiveness of LLMs with respect to code generation has been investigated for a specific languages such as Python~\cite{Tambon2024} and Verilog~\cite{Thakur2023}, as well as for particular programming paradigms, such as the use of classes in an object-oriented setting~\cite{XueyingDu_C_2023}.
Xu et al.\ systematically evaluated multilingual LLMs for code correctness and perplexity~\cite{xu2022}. Moradi et al.\ compared Copilot's solutions to human-generated code on algorithmic problems~\cite{dakhel2023}.
The focus of CCTest~\cite{li2023} is on improving LLM-based code completion by ensuring structural consistency using Levenshtein edit distance~\cite{levenshtein1966}, and detecting errors through mutations that maintain consistency of program structure.
Wong et al.\ assessed Copilot's code quality by formally verifying that generated code meets predefined specifications~\cite{wong2022}. 
Rajan et al.\ proposed KONTEST to detect inconsistencies in LLM outputs using knowledge graph-based test cases and metamorphic and ontological oracles~\cite{Rajan2024}. 
Dozono et al.\ evaluated LLMs for detecting common weaknesses and introduced CODEGUARDIAN to enhance accuracy and speed in VS Code~\cite{Dozono2024}.
Recent works have focused on the importance of fine-tuning in LLM performance~\cite{Zhiqiang_Yuan2023},
and on
evaluating self-consistency of LLMs in code generation and comprehension tasks~\cite{Marcus2024}.

In contrast to these works, which mainly focus on the absolute performance of LLMs on specific code generation tasks, our approach evaluates LLM code generation capabilities across \emph{neighbourhoods} of related question instances,
allowing the identification of discontinuities in reasoning ability.
This can offer insights into how LLMs handle a variety of strongly-related tasks in a given problem space,
which is under-explored by these prior works. 
Additionally, unlike most previous approaches that rely on manually crafted test suites, we automate testing by combining fixed test suites with fuzzing as complementary techniques~\cite{ciupa2008}. 

\ourpara{Robustness} Several studies have investigated LLM robustness to syntactic variations that preserve semantics,
e.g.\
by modifying problem descriptions without altering semantics~\cite{shirafuji2023,anand2021},
perturbing prompts (with the finding that slight perturbations can significantly impact model performance)~\cite{wang2022,doderlein2022}, and changing method names~\cite{Guang_Yang2024}.
Various works have focused on enhancing robustness:
CLAWSAT utilises contrastive learning with adversarial views and staggered adversarial training for this purpose~\cite{Jinghan_Jia2023},
the CoTR framework~\cite{guangYang2023} defends code translation models against adversarial attacks through syntactic transformations and data augmentation with semantically equivalent code examples,
CodeBERT-Attack highlights vulnerabilities and suggests adversarial training examples for model improvement~\cite{CodeBERT_Attack2024}, the CODA framework aims to enhance model robustness by generating adversarial examples from semantically similar inputs~\cite{Zhao_Tian2023}, and the CARROT framework focuses on robustness detection, measurement, and enhancement in the context of code-focused LLMs~\cite{Huangzhao_Zhang2022}.
Numerous other approaches focus on the problem of identifying a lack of robustness in models~\cite{yang2022,mingYan2023,Huangzhao_Zhang2020,Junkai_chen2024,Zhengran_Zeng2022,yefet2020,Srikant2021,Moshi_Wei2023}.

Among these works, the study by Shirafuji et al.~\cite{shirafuji2023} is the most closely related to our research. However, there is a key difference in focus. Their study examines how syntactic modifications, such as altering variable names or rephrasing prompts, influence the correctness and quality of generated code while maintaining the underlying task. In contrast, our research explores how LLMs perform when faced with a question neighbourhood—a set of \textit{semantically similar but distinct tasks}. By leveraging parameterised question templates, we systematically investigate the models’ ability to generalise and identify gaps in their performance.

\section{Conclusions and Future Work}\label{sec:conclusion}
We have introduced a new method for assessing the correctness and robustness of LLMs with respect to code generation, based on the notion \emph{question neighbourhoods}.
Being able to assess the performance of an LLM across a question neighbourhood makes it possible not only to identify specific problem instances that an LLM cannot solve, but to identify gaps in an LLM's ability to perform general reasoning in a particular problem space.
We have put this into practice via \bmk{}, the first benchmark to systematically evaluate code-generating LLMs using question neighbourhoods. Experiments with five models showed that \gfour{} consistently outperformed the other evaluated models, but that all models demonstrated a lack of robustness in certain question neighbourhoods. Lowering the temperature to zero improved correctness scores (except for \com{}) and reduced error diversity.

Interesting avenues for future research include assessing the impact of quantisation on LLM performance, and developing \bmk{}-like benchmarks for code-infilling models.

\section{Data Availability Statement}
The source code for \bmk{}, all question and oracle templates, together with all results, are available in the Zenodo repository \cite{Honarvar2025}.

\clearpage

\bibliographystyle{IEEEtran}
\bibliography{ref}

\end{document}

%% file: question_classes.tex
\begin{center}
\renewcommand{\arraystretch}{1.1}
\setlength{\tabcolsep}{4pt}
\resizebox{\columnwidth}{!}{
\begin{tabular}{|l|l|r|}
\hline
\textbf{Problem Group} & \textbf{Problem Subgroup} & \textbf{Question Count} \\
\arrayrulecolor{black}\hline

List Manipulation & Total & 40 \\
\arrayrulecolor{black}\cdashline{2-3} 
& Slicing & 21 \\
& Indexing & 14 \\
& Filtering & 16 \\ 
& Element-based Operations & 7 \\
& Summation & 6 \\
& Sorting/Order-based Operations & 6 \\
& Element Insertion & 2 \\
& Count elements & 2 \\
& Circular Lists & 1 \\
\arrayrulecolor{black}\hline

String Manipulation & Total & 16 \\
\arrayrulecolor{black}\cdashline{2-3} 
& Character Insertion & 2 \\
& Character Removal & 3 \\
& Substring/Character Extraction & 4 \\
& Palindrome Operations & 4 \\
& Anagram Detection & 2 \\
& Sorting & 2 \\
\arrayrulecolor{black}\hline

Set Manipulation & Total & 9 \\
\arrayrulecolor{black}\cdashline{2-3} 
& Add Elements & 6 \\
& Subset/Superset Operation & 1 \\
& Counting Subsets & 1 \\
& Union & 1 \\
\arrayrulecolor{black}\hline

Searching & Total & 36 \\
\arrayrulecolor{black}\cdashline{2-3} 
& Linear Search & 12 \\
& Binary Search & 8 \\
& Index-based Search & 10 \\
& String Search & 6 \\
\arrayrulecolor{black}\hline

Copying & Total & 10 \\
\arrayrulecolor{black}\cdashline{2-3} 
& Deep Copy & 4 \\
& Shallow Copy & 3 \\
& Copy Sublist & 3 \\
\arrayrulecolor{black}\hline

Mathematical Problems & Total & 31 \\
\arrayrulecolor{black}\cdashline{2-3} 
& Arithmetic Operations & 7 \\
& Factorial Calculations & 5 \\
& Prime Checking & 4 \\
& Composite Checking & 3 \\
& Factorisation & 4 \\
& Special Sequences & 3 \\
& Combinatorial Problems & 5 \\
\arrayrulecolor{black}\hline

\end{tabular}
}
\end{center}

%% file: failure_reasons.tex
\begin{center}
\footnotesize
\resizebox{\textwidth}{!}{
\begin{tabular}{| c | rr | rr | rr | rr | rr |}
\hline
& \multicolumn{2}{c|}{\textbf{\clsev{}}} & \multicolumn{2}{c|}{\textbf{\clthir{}}} & \multicolumn{2}{c|}{\textbf{\com}} & \multicolumn{2}{c|}{\textbf{\gt{}}} & \multicolumn{2}{c|}{\textbf{\gfour}}\\
\textbf{Categories} & \multicolumn{1}{r}{\tzerotable{}} & \multicolumn{1}{r|}{\tdtable{}} & \multicolumn{1}{r}{\tzerotable{}} & \multicolumn{1}{r|}{\tdtable{}} & \multicolumn{1}{r}{\tzerotable{}} & \multicolumn{1}{r|}{\tdtable{}} & \multicolumn{1}{r}{\tzerotable{}} & \multicolumn{1}{r|}{\tdtable{}} & \multicolumn{1}{r}{\tzerotable{}} & \multicolumn{1}{r |}{\tdtable{}}\\\hline
no function &0.00\% &0.00\% &0.00\% &0.00\% &0.00\% &0.45\% &0.00\% &0.13\% &0.00\% &0.00\% \\
wrong function name &0.00\% &0.02\% &0.00\% &0.14\% &0.00\% &0.02\% &0.00\% &0.07\% &0.00\% &0.24\% \\
wrong count of arguments &3.73\% &4.04\% &0.82\% &2.69\% &0.00\% &0.02\% &0.00\% &0.00\% &0.00\% &0.00\% \\
syntax error &4.90\% &11.72\% &0.47\% &0.83\% &1.14\% &1.96\% &0.00\% &0.04\% &0.00\% &0.07\% \\
static type error &3.57\% &9.34\% &10.28\% &9.29\% &7.68\% &5.88\% &0.80\% &0.91\% &0.03\% &0.05\% \\
resource exhaustion &0.63\% &0.63\% &0.02\% &0.12\% &0.04\% &0.37\% &2.26\% &1.22\% &0.93\% &0.91\% \\
runtime error &7.65\% &8.10\% &4.74\% &6.85\% &10.81\% &13.40\% &1.95\% &2.98\% &0.56\% &1.04\% \\
assertion error &45.65\% &36.35\% &44.09\% &43.90\% &66.91\% &64.35\% &27.15\% &26.65\% &9.31\% &10.89\% \\
fuzzing failure &4.92\% &4.98\% &6.85\% &6.93\% &7.86\% &7.28\% &6.17\% &6.75\% &4.38\% &4.66\% \\
\hline
passed &28.95\% &24.81\% &32.73\% &29.25\% &5.57\% &6.27\% &61.67\% &61.26\% &84.79\% &82.15\% \\
\hline
\end{tabular}
}
\end{center}